\documentstyle[12pt]{article}
\begin{document}

\begin{flushright}
PITHA 94/52\\
hep-ph/9411374\\
November, 1994
\end{flushright}
\vspace*{2cm}
\LARGE\centerline{Rare $K$ Decays \footnote{Invited talk at the Third
Workshop on High Energy Particle Physics (WHEPP 3), Madras, Jan. 10 -
Jan. 22, 1994. To appear in Proceedings.}}
\vspace*{1cm}
\large\centerline{L. M. Sehgal}
\centerline{III. Physikalisches Institut (A), RWTH Aachen,}
\centerline{D-52074 Aachen, Germany}
\normalsize
\vspace*{2cm}

\indent The rare decays of the $K$ meson have had a long tradition as a
laboratory for testing the symmetry properties of the weak interactions,
and the manner in which these symmetries are broken by higher order
effects. Present--day interest is focussed on decays that are suppressed
by $CP$--symmetry or GIM symmetry. Such decays, in the standard theory,
are sensitive to effects of the virtual top quark, and could also reveal
new interactions transcending the standard model. In addition, the
radiative decays of the $K$ meson have become a useful testing--ground
for effective Lagrangians describing the low energy interactions of
pions, kaons and photons.\\
\indent This talk is a selective review of some rare $K$ processes.
For a more comprehensive discussion, we refer to the reviews listed in [1].
\vspace*{2cm}

\section{Charged Current Rarities}
An example of a decay that is rare, yet allowed in the lowest order, is
the $\Delta S=0$ transition $K^0\to K^+e^-\bar{\nu_e}$ (Fig. 1a). This is
closely analogous to pion $\beta $-decay, $\pi^-\to\pi^0e^-\bar{\nu_e}$.
Conservation of the vector current $\bar u\gamma_{\mu}d$ dictates that
the relevant matrix element is $\langle K^0|I_-|K^+\rangle = 1$ (in
analogy to $\langle \pi^0|I_-|\pi^+\rangle = \sqrt 2$) and the predicted
branching ratio is
\begin{eqnarray}
B(K_L\to K^+e^-\nu_e) = 3\times 10^{-9}.
\end{eqnarray}
\indent A curious analogue of the $\Delta S = \Delta Q$ decay $K^+\to
\pi^0e^+\nu_e$ is
the hypothetical transition $K^+\to ``\eta\mbox{''}e^+\nu_e$. The latter
is not observable as a real decay process, since $m_{\eta}>m_K$. The
matrix elements for $K^+\to\eta $ and $K^+\to\pi^0$ are $-1/\sqrt 6$ and
$1/\sqrt 2$, respectively. A possible way to probe the $K^+\to\eta $
coupling is via the decay
\begin{eqnarray}
K^+\to\gamma \gamma e^+\nu_e.
\end{eqnarray}
Some of the relevant diagrams are shown in Fig. 1b. An estimate of this
decay would be of interest.\\
\indent In the second order of weak interactions, it is possible to obtain
a $\Delta S = -\Delta Q$ transition $K^0\to\pi^+e^-\bar{\nu_e}$ (Fig. 1c).
A model based on the parity--conserving vertices $K^0-\pi^0$ and $K^+-
\pi^+$, combined with the $\Delta S = \Delta Q$ transitions $\pi^0\to
\pi^+l^-\bar{\nu_l}$ and $K^0\to K^+l^-\bar{\nu_l}$ gives \cite{2}
\begin{eqnarray}
\frac{\Gamma (\Delta S = -\Delta Q)}{\Gamma (\Delta S = +\Delta Q)}=0.5
\times 10^{-12}\qquad (K_{e3},K_{\mu 3}).
\end{eqnarray}
An alternative mechanism, combining the parity--violating decay $K^0\to
\pi^+\pi^-$ with the transition $\pi^-\to\mu^-\bar{\nu_{\mu}}$, yields
\cite{3}
\begin{eqnarray}
\frac{\Gamma (\Delta S = -\Delta Q)}{\Gamma (\Delta S = +\Delta Q)}=1.1
\times 10^{-12}
\qquad (K_{\mu 3} \mbox{ only}).
\end{eqnarray}
This second mechanism produces a $\Delta S = -\Delta Q$ amplitude $K^0
\to\pi^+\mu^-\nu_{\mu}$ with a helicity structure and pion energy
spectrum that is quite different from that in the $\Delta S = \Delta Q$
transition $\bar{K^0}\to\pi^+\mu^-\bar{\nu_{\mu}}$. To the extent that
both mechanisms are possible, the $e/\mu$ ratio in $\Delta S = -\Delta Q$
transitions will be different from that in $\Delta S = \Delta Q$ decays.

\section{Neutral Current Decay into $l^+l^-$ Pairs}
\subsection{Decay $K_L\to\mu^+\mu^-$}
The decay $K_L\to\mu^+\mu^-$ has a well--known unitarity bound \cite{4}
\begin{eqnarray}
\frac{\Gamma (K_L\to\mu^+\mu^-)}{\Gamma (K_L\to 2\gamma )}\ge
\frac{\alpha^2}{2\beta}
\frac{m_{\mu}^2}{m_K^2}\Bigl (\ln\frac{1+\beta }{1-\beta}\Bigr )^2,
\qquad\beta = \sqrt{1-4m_{\mu}^2/m_K^2}
\end{eqnarray}
associated with the $2\gamma $ intermediate state (Fig. 2a).
The corresponding branching ratio is
\begin{eqnarray}
B_{abs}=(6.8\pm 0.3)\times 10^{-9}.
\end{eqnarray}
If one includes the dispersive part of the $2\gamma $ amplitude, as well
as the real amplitude induced by the short--distance interaction $s\bar d
\to\mu^+\mu^-$ (box and penguin diagrams, Fig. 2b), the full
branching ratio is
\begin{eqnarray}
B(K_L\to\mu^+\mu^-)=B_{abs}+|\sqrt{B_{disp}}\pm
\sqrt{B_{short-dist.}}|^2.
\end{eqnarray}
\indent A theoretical estimate of $B_{disp}$ requires a model for the
$K_L\to\gamma^*\gamma^*$ form factor. Vector--meson--dominance models
\cite{5}, similar to
those used for $\pi^0 ,\eta\to\gamma^*\gamma^*$, tend to give $B_{disp}/
B_{abs}\le 0.1$ (to be compared with the experimental rate of
$\eta\to\mu^+\mu^-$,
$B(\eta\to\mu^+\mu^-)=(1.3\pm 0.1)\times B_{abs}(\eta\to\mu^+\mu^-)$).
A model for $K_L\to\gamma^*\gamma$ \cite{6}, that describes the observed
spectrum of Dalitz pairs in $K_L\to\gamma e^+e^-$ \cite{7}, yields an even
smaller value of $B_{disp}$.\\
\indent The short--distance branching ratio has been calculated to be
\cite{8,9}
\begin{eqnarray}
B_{short-dist.}(K_L\to\mu^+\mu^-)=6.4\times 10^{-3}\nonumber\\
\cdot\Bigl [P_0(K_L\to 2\mu)+\mbox{Re}\lambda_tY(x_t)\Bigr ]^2
\end{eqnarray}
where
\begin{eqnarray}
Y(x_t)=\frac{x_t}{8}\Bigl [\frac{x_t-4}{x_t-1}+\frac{3x_t}{(x_t-1)^2}
\ln x_t\Bigr ], \qquad x_t=\frac{m_t^2}{m_W^2},
\end{eqnarray}
and
\begin{eqnarray}
\lambda_t=V^*_{ts}V_{td}=-A^2\lambda^5(1-\rho -i\eta ).
\end{eqnarray}
The term $P_0(K_L\to 2\mu)$ denotes a residual charm quark contribution,
estimated to be $-0.75\times 10^{-4}$, for $m_c=1.4$ GeV,
$\Lambda = 200 $ MeV.\\
\indent The experimental branching ratio $B_{exp}(K_L\to\mu^+\mu^-)=
(7.4\pm 0.4)\times 10^{-9}$ \cite{10}, is only slightly in excess of
the unitarity bound (Eq. (6)), and  can constrain the parameter
$\mbox{Re}\lambda_t=-A^2\lambda^5(1-\rho )$ appearing in the
short--distance contribution. At present, the uncertainty in the
$2\gamma$ dispersive rate $B_{disp}$ limits the efficacy of this decay
in providing a stringent constraint on $\rho$ \cite{11}.

\subsection{Decay $K_L\to\pi^0l^+l^-$}
In the one--photon exchange approximation, $CP$--symmetry forbids the
decay $K_2\to\pi^0l^+l^-$ but allows the decay $K_1\to\pi^0l^+l^-$
\cite{12}. A $CP$--conserving amplitude for $K_2\to\pi^0l^+l^-$ is
possible through a $2\gamma$ intermediate state (Fig. 2a).
The $2\gamma $--induced
amplitude is proportional to $m_l$ if the two photons have $J=0$;
however, dynamical models of $K_2\to\pi^0\gamma\gamma$ permit also a
$J=2$ component which yields an amplitude for $K_2\to\pi^0l^+l^-$
unsuppressed by a factor $m_l$ \cite{13}. It follows that the
amplitude for $K_L\to\pi^0l^+l^-$ will contain three components
\begin{eqnarray}
A=\alpha^2 A_{2\gamma}+\alpha\varepsilon A_{1\gamma}+
\eta\lambda^4 A_{sd},
\end{eqnarray}
where the second and third pieces denote amplitudes associated with
indirect and direct (short--distance) $CP$--violation. The factors
$\alpha^2$, $\varepsilon\alpha$ and $\eta\lambda^4$ express the orders
of magnitude of these components, and underscore the
fact that they are, a priori, comparable in size.\\
\indent The $CP$--conserving amplitude for $K_2\to\pi^0l^+l^-$
involving a $2\gamma $ intermediate state has the form \cite{14}
\begin{eqnarray}
A(K_2(p)\to\pi^0 (p')+l^-(k)+l^+(k'))_{2\gamma}=-\frac{\alpha}{16}
\nonumber\\
\cdot\bar u(k)\Bigl [
\frac{G_{eff}}{M_{\rho}^2}F_1p\!\!/+F_2\Bigr ]v(k'),
\end{eqnarray}
where $F_2$ is proportional to $m_l$, and is negligible for $K_2\to
\pi^0e^+e^-$. The coefficient $G_{eff}/M_{\rho}^2$ is an effective
coupling constant appearing in the $K_2\to\pi^0\gamma\gamma$ amplitude,
parametrized as
\begin{eqnarray}
A(K_2(p)&\to&\pi^0(p')\gamma (q)\gamma '(q'))=2
\frac{G_{eff}}{M_{\rho}^2}\Bigl [
(\epsilon\cdot\epsilon ')(q\cdot p)(q'\cdot p)\nonumber\\
&+&(q\cdot q')(\epsilon\cdot p)(\epsilon '\cdot p)
-(\epsilon\cdot p)(\epsilon '\cdot q)(q'\cdot p)\nonumber\\
&-&(\epsilon\cdot q')(q\cdot p)(\epsilon '\cdot p)\Bigr ]+\cdots
\end{eqnarray}
where the ellipsis denote terms that are ineffective for $K_L\to
\pi^0e^+e^-$. An analysis of the data on $K_L\to\pi^0\gamma\gamma$
\cite{15} suggests $G_{eff}\sim 0.15\times
10^{-7}m_K^{-2}$ \cite{14}. The form factor $F_1$ has an absorptive part
\begin{eqnarray}
\mbox{Im} F_1&=&-\frac{\Delta}{\beta^2}\Bigl [\frac{2}{3}+
\frac{2}{\beta^2}-(\frac{1}{\beta^2}-\beta^2)\frac{1}{\beta}
\ln\frac{1+\beta}{1-\beta}\Bigr ]\nonumber\\
&\stackrel{\beta \to 1}{\to}&\frac{8}{3}\Delta ,
\end{eqnarray}
where $\Delta =-2p\cdot (k-k')$, $\beta = \sqrt{1-4m_l^2/s}$,
$s=(k+k')^2$. The corresponding $CP$--conserving branching ratio is
\begin{eqnarray}
B_{CPC}(K_L\to\pi^0e^+e^-)=1.7\times 10^{-12}\Bigl (
\frac{G_{eff}m_K^2}{0.15\times 10^{-7}}\Bigr )^2(1+\rho_{disp}),
\end{eqnarray}
$\rho_{disp}$ being the dispersive part, estimated in \cite{14} to be
$1.5$.\\
\indent The result (15) has to be compared with the $CP$--violating
rate \cite{16}
\begin{eqnarray}
B_{CPV}(K_L\to\pi^0e^+e^-)=\Bigl [|0.76re^{i\pi/4}+i\tilde{C_V}
\frac{\eta}{0.4}|^2\nonumber\\
+|\tilde{C_A}\frac{\eta}{0.4}|^2\Bigr ]\times 10^{-11},
\end{eqnarray}
where $\tilde{C_V}$ and $\tilde{C_A}$ are coupling constants describing
the short--distance interaction $s\bar d\to e^+e^-$, with numerical
values $\tilde{C_V}=-0.6$, $\tilde{C_A}=0.7$ for $m_t = 170$ GeV. The
parameter $r$, characterising the indirect $CP$--violating amplitude,
is defined as
\begin{eqnarray}
r=\Bigl [\frac{\Gamma(K_1\to\pi^0e^+e^-)}{\Gamma (K^+\to\pi^+e^+e^-)}
\Bigr ]^{1/2}.
\end{eqnarray}
For $\eta = 0.4$, the $CP$--violating branching ratio $B_{CPV}$ is
$1.2\times  10^{-11}(2.1\times 10^{-11})$ for $r=+1(-1)$. Models with
much larger and much smaller values of $r$ are possible.
These estimates suggest that the $CP$--violating rate $B_{CPV}$ is
somewhat higher than the $CP$--conserving rate $B_{CPC}$. A precise
measurement of the rate and spectrum of $K_2\to\pi^0\gamma\gamma$ would
help to sharpen the estimate of $B_{CPC}$, while a measurement of $r$
is needed to predict $B_{CPV}$. Expectations for the decay
$K_L\to\pi^0\mu^+\mu^-$, as well as an estimate of the $CP$--violating
$l^+l^-$ energy asymmetry are given in \cite{14}. Present experimental
limits (from the E799 experiment at Fermilab) are
$B(K_L\to\pi^0e^+e^-)<4.3\times 10^{-9}$,
$B(K_L\to\pi^0\mu^+\mu^-)<5.1\times 10^{-9}$ \cite{17}.

\subsection{Decay $K_L\to\pi^+\pi^-e^+e^-$}
A study of the photon spectrum in the decay $K_L\to\pi^+\pi^-\gamma $
shows two clear components \cite{18}: (i) bremsstrahlung from the
$CP$--violating decay $K_L\to\pi^+\pi^-$, and (ii) direct photon
emission of magnetic dipole nature from the $CP$--conserving decay
$K_2\to\pi^+\pi^-\gamma$. The simultaneous presence
of bremsstrahlung and $M1$ amplitudes implies that the photon in the
decay $K_L\to\pi^+\pi^-\gamma$ has a $CP$--violating circular
polarization. The conversion process $K_L\to\pi^+\pi^-e^+e^-$ may
be viewed as a means of probing this polarization, by studying the
correlation of the $e^+e^-$ plane relative  to the
$\pi^+\pi^-$ plane.\\
\indent An analysis based on the amplitude
\begin{eqnarray}
A(K_L\to\pi^+(p_+)\pi^-(p_-)e^+(k_+)e^-(k_-))=e|f_S|\Bigl [g_{Br}(
\frac{p_+^{\mu}}{p_+\cdot k}-\frac{p_-^{\mu}}{p_-\cdot k})\nonumber\\
+g_{M1}\varepsilon_{\mu\nu\rho\sigma}
k^{\nu}p_+^{\rho}p_-^{\sigma}\Bigr ]\frac{e}{k^2}\bar u(k_-)
\gamma_{\mu}v(k_+)
\end{eqnarray}
was carried out in \cite{19}. Here $k=k_++k_-$, $f_S$ is the
amplitude of $K_S\to\pi^+\pi^-$, and $g_{Br}$ and $g_{M1}$ are
given empirically by $g_{Br}=\eta_{+-}e^{i\delta_0 (m_K^2)}$,
$g_{M1}=i(0.76)e^{i\delta_1 (s_{\pi})}$, $\delta_{0,1}$ being the
$s$-- and $p$--wave $\pi\pi$ phase shifts. A significant
$CP$--violating
asymmetry was found in the $\Phi$--distribution of the process,
$\Phi$ being the angle between the $e^+e^-$ and $\pi^+\pi^-$ planes:
\begin{eqnarray}
A&=&\displaystyle\frac{\displaystyle\int\limits_0^{\pi /2}d\Phi
\displaystyle\frac{d\Gamma}{d\Phi}-
\displaystyle \int\limits_{\pi /2}^{\pi}d\Phi
\displaystyle\frac{d\Gamma}{d\Phi}}{ \displaystyle\int
\limits_0^{\pi }d\Phi\displaystyle\frac{d\Gamma}{d\Phi}}\nonumber\\
&=&15\% \sin (\Phi_{+-}+\delta_0 (m_K^2)-\bar{\delta_1})\nonumber\\
&\approx &14\%,
\end{eqnarray}
where $\bar{\delta_1}$ denotes an average $p$--wave phase
($\sim 10^0$). This analysis was extended in \cite{20} to include
short--distance $CP$--violation, contained in the effective
Hamiltonian for $s\bar d\to e^+e^-$. These direct $CP$--violating
effects were found to be very small ($<10^{-3}$). The branching
ratio of $K_L\to\pi^+\pi^-e^+e^-$ is predicted to be
$3\times 10^{-7}$, so that the large asymmetry given in Eq. (19)
may well be accessible in the next round of experiments.

\subsection{Decay $K^+\to\pi^+l^+l^-$}
The decays $K^+\to\pi^+l^+l^-$ are dominated by one--photon exchange,
and are principally of interest as tests of the $K\pi\gamma^*$ vertex.
In chiral pertubation theory, this vertex is calculable in terms of
the effective Lagrangian ${\cal L}_{eff}(\pi ,K,\gamma )$,
and the matrix element has the form \cite{21}
\begin{eqnarray}
\label{24_1}
A(K^+(k)\to\pi(p)+l^++l^-)=\frac{\alpha G_8}{4\pi}C^+(z)
\bar u(k\!\!/+p\!\!/)v,\nonumber\\
 z=(k-p)^2/m_K^2,
\end{eqnarray}
where $G_8=G_F/\sqrt 2 V_{ud}V^*_{us}g_8$, $g_8=5.1$. The function
$C^+(z)$ is known, up to an additive constant $w_+$. In terms of
this constant, the branching ratios are calculated to be
\begin{eqnarray}
B(K^+\to\pi^+e^+e^-)&=&(3.15-21.1w_++36.1w_+^2)\times 10^{-8},\
\nonumber\\
B(K^+\to\pi^+\mu^+\mu^-)&=&(3.93-32.7w_++70.5w^2_+)\times 10^{-9}.
\end{eqnarray}
The recent measurement \cite{22} $B(K^+\to\pi^+e^+e^-)=
(2.99\pm 0.22)\times 10^{-7}$ implies $w_+=0.89^{+0.26}_{-0.14}$,
and leads to the prediction $B(K^+\to\pi^+\mu^+\mu^-)=3.07
\times 10^{-8}$.\\
\indent Refinements to the matrix element (\ref{24_1}) occur
if one takes into account the short--distance interaction
$s\bar d\to l^+l^-$ \cite{23}. One
interesting aspect of these corrections is the addition to the
matrix element (\ref{24_1}) of a parity--violating term of the
form
\begin{eqnarray}
\label{24_4}
A_{sd}(K^+(k)\to\pi(p)+l^++l^-)=\frac{G_F\alpha}{\sqrt 2}V_{us}
\Bigl [B(k+p)^{\mu}+C(k-p)^{\mu}\Bigr ]\nonumber\\
\bar u\gamma_{\mu}\gamma_5v.
\end{eqnarray}
The short--distance interaction gives $B=f_+\xi$, $C=f_-\xi$,
where
\begin{eqnarray}
\xi = -1.4\times 10^{-4}-\frac{Y(x_t)}{2\pi\sin^2\theta_W}A^2
\lambda^4(1-\rho-i\eta ),
\end{eqnarray}
$f_+$, $f_-$ being the two form factors of $K_{l3}$ decay
($f_+\approx 1$, $f_-\approx 0$). Interference with the
leading one--photon exchange amplitude (\ref{24_1}) gives rise
to a parity--violating longitudinal
polarization of the $\mu^+$ \cite{23}
\begin{eqnarray}
|A_{LR}|\equiv |\frac{\Gamma_L-\Gamma_R}{\Gamma_L+\Gamma_R}|=
|2.3\mbox{Re}\xi |,
\end{eqnarray}
thus providing a way to determine the parameter $\rho$.\\
\indent Another consequence \cite{24} of the short--distance
amplitude (\ref{24_4}) (which we write as $F_A\bar u(p\!\!/+k\!\!/)
\gamma_5v$) is that its interference with the dominant one--photon
amplitude (\ref{24_1}) (which we write as $F_V\bar
u(p\!\!/+k\!\!/)v$) produces a $T$--odd, $P$--odd term in the decay
rate, of the form $\mbox{Im}(F_VF_A^*)(\vec s_+\times\vec s_-)\cdot
\vec p$, where $\vec s_+$ and $\vec s_-$ are spin vectors of the
$\mu^+$ and $\mu^-$ in the $K$ rest frame, and $\vec p$ is the
$\mu^+$ momentum. Such a term can probe the $CP$-violating parameter
$\eta$ of the short--distance interaction. Detection of this term,
however, requires measurement of a correlation between
the spins of both $\mu^+$ and $\mu^-$, a difficult task.\\
\indent An important challenge for the low energy effective
Lagrangian ${\cal L}_{eff}
(K,\pi ,\gamma)$ is to produce a reliable prediction for the decay
$K_1\to\pi^0l^+l^-$. In chiral pertubation theory, the matrix element
is determined up to an unknown constant
$w_S$ \cite{21}. As discussed in Section 2.2, information on this
decay mode is important for determining the magnitude of indirect
$CP$--violation in $K_L\to\pi^0l^+l^-$.

\section{Neutral Current Decays into $\nu\bar{\nu}$ Pairs}
\subsection{Decay $K^+\to\pi^+\nu\bar{\nu}$}
The decay $K^+\to\pi^+\nu\bar{\nu}$ is a short--distance dominated
reaction, determined by the box and penguin graphs shown in Fig. 3a.
The branching ratio is predicted to be \cite{25}
\begin{eqnarray}
B(K^+\to\pi^+\nu_l\bar{\nu_l})=5.9\times 10^{-5}|P_0(K^+\to\pi^+
\nu_l\bar{\nu_l})+\lambda_tX(x_t)|^2,
\end{eqnarray}
where
\begin{eqnarray}
P_0(K^+\to\pi^+\nu_l\bar{\nu_l})=\left\{
    \begin{array}{l@{\quad}l}
    -2.5\times 10^{-4}&\mbox{for}\qquad\nu_e,\nu_{\mu}\\
    -1.7\times 10^{-4}&\mbox{for}\qquad\nu_{\tau}
    \end{array} \right.
\end{eqnarray}
and
\begin{eqnarray}
X(x_t)=\frac{x_t}{8}\Bigl [\frac{x_t+2}{x_t-1}+
\frac{3x_t-6}{(x_t-1)^2}\ln x_t\Bigr ].
\end{eqnarray}
Summing over all three neutrino flavours, one obtains a typical
value $B(K^+\to\pi^+\nu\bar{\nu})=1.3\times 10^{-10}$ for $\rho = 0$,
$\eta = 0.4$, with a possible range $(0.5-5)\times 10^{-10}$ for
the presently allowed domain of $(\rho ,\eta )$. The present
experimental limit is $5.2\times 10^{-9}$ (AGS E787) \cite{26} and
a sensitivity of $10^{-10}/\mbox{event}$ is within reach.\\
\indent The long--distance contributions to
$K^+\to\pi^+\nu\bar{\nu}$, symbolised by the diagrams in Fig. 3b,
were calculated in \cite{27}. In particular, the hadronic contribution
to the $K^+\pi^+Z$ vertex was obtained using current algebra arguments.
It was concluded that these effects are three orders of magnitude
smaller than the short--distance contribution. A more recent
calculation \cite{28}, using chiral pertubation theory, has found
a very similar result. (For additional remarks, see \cite{29}).

\subsection{Decay $K_L\to\pi^0\nu\bar{\nu}$}
Finally, an example of a short--distance dominated process, which is
at the same time purely $CP$--violating, is the decay
$K_2\to\pi^0\nu\bar{\nu}$ \cite{30}. Its branching ratio is given
by \cite{31}
\begin{eqnarray}
B(K_L\to\pi^0\nu\bar{\nu})&=&7.32\times 10^{-4}\Bigl [\mbox{Im}
\lambda_t\Bigr ]^2X^2(x_t)\nonumber\\
&=&1.94\times 10^{-10}\Bigl [\eta^2A^4\Bigr ]X^2(x_t).
\end{eqnarray}
While experimentally remote (the present limit is
$B(K_L\to\pi^0\nu\bar{\nu})<5.8\times 10^{-5}$ \cite{32}), this
reaction is an interesting example of a process that directly measures
the $CP$--violating paramter $\eta $, with essentially no hadronic
uncertainties.

\section{Radiative Decays}
The radiative decays of the $K$ mesons, such as
$K_{L,S}\to\gamma\gamma$, $K_L\to\pi^0\gamma\gamma$,
$K_L\to\pi^+\pi^-\gamma$, $K^+\to\pi^+\pi^0\gamma $,
$K^+\to\pi^+\gamma\gamma$, $K_L\to\pi^0\pi^0\gamma\gamma$ ,$K_{L,S}\to
\pi^0\pi^0\gamma$ etc. are a source of abundant grist for models,
such as chiral pertubation theory \cite{33}, that attempt to describe
the low energy interactions of pions, kaons and photons. The interplay
of chiral symmetry, $CP$--symmetry and gauge invariance, combined
with the weak nonleptonic $\Delta I=1/2$ rule, produces interesting
patterns and hierarchies amongst the various channels. We will limit
our remarks here to two reactions,
\begin{eqnarray}
K_{L,S}&\to &\pi^0\pi^0\gamma ,\nonumber\\
K_{L,S}&\to &3\gamma
\end{eqnarray}
that have the piquant feature of being {\it quadrupole } transitions.

\subsection{Decay $K_{L,S}\to \pi^0\pi^0\gamma $}
Gauge invariance implies that the $\pi^0\pi^0$ system in
$K_{L,S}\to \pi^0\pi^0\gamma $ cannot have $J=0$ (since that would
amount to a $0\to 0$ radiative transition). Bose statistics implies
that the $\pi^0\pi^0$ pair cannot have $J=1$ (since such a state
would not be symmetric under exchange of the two $\pi^0$'s). It
follows that the $\pi^0\pi^0$ state has at least two units of angular
momentum, and the associated photon corresponds
to quadrupole radiation \cite{34}.\\
\indent $CP$--invariance implies that the decays
$K_L\to\pi^0\pi^0\gamma$ and $K_S\to\pi^0\pi^0\gamma$ are $E2$ and
$M2$ transitions, respectively, with matrix elements
\begin{eqnarray}
A(K_L\to\pi^0(p_1)\pi^0(p_2)\gamma (k))&=&\frac{g_{E2}}{m_K^3}
\frac{(p_1-p_2)\cdot k}{\Lambda^2}\Bigl [(\epsilon\cdot p_1)
(k\cdot p_2)\nonumber\\
&&-(\epsilon\cdot p_2)(k\cdot p_1)\Bigr ],\nonumber\\
A(K_S\to\pi^0(p_1)\pi^0(p_2)\gamma (k))&=&\frac{g_{M2}}{m_K^3}
\frac{(p_1-p_2)\cdot k}{\Lambda^2}\varepsilon_{\mu\nu\rho\sigma}
\epsilon^{\mu}k^{\nu}p_1^{\rho}p_2^{\sigma}.
\end{eqnarray}
A qualitative estimate of the decay rates may be obtained by making a
comparison with the measured $M1$ transition $K_L\to\pi^+\pi^-\gamma$
\cite{16}, parametrized as
\begin{eqnarray}
A(K_L\to\pi^+(p_+)\pi^-(p_-)\gamma (k))=\frac{g_{M1}}{m_K^3}
\varepsilon_{\mu\nu\rho\sigma}\epsilon^{\mu}k^{\nu}
p_+^{\rho}p_-^{\sigma}.
\end{eqnarray}
Assuming that the dimensionless couplings $g_{M1}$, $g_{M2}$
and $g_{E2}$ are similar in magnitude, and that the reaction
``radius'' $1/\Lambda $ is of order $1/m_{\rho}$, we obtain \cite{34}
\begin{eqnarray}
B(K_L\to\pi^0\pi^0\gamma )&\approx &1.0\times 10^{-8},\nonumber\\
B(K_S\to\pi^0\pi^0\gamma )&\approx &1.7\times 10^{-11}.
\end{eqnarray}
\indent In chiral pertubation theory, the reaction
$K_L\to\pi^0\pi^0\gamma$ occurs only in order $p^6$ in the chiral
expansion \cite{35}. On the other hand the reaction
$K_L\to\pi^0\pi^0\gamma^*$, with a virtual photon, is possible
in order $p^4$. It follows that a study of the Dalitz pair process
$K_L\to\pi^0\pi^0e^+e^-$ could show two components, one associated
with $K_L\to\pi^0\pi^0\gamma$ ($E2$; $O(p^6)$), which is strongly
peaked at low $e^+e^-$ masses, and one associated with
$K_L\to\pi^0\pi^0\gamma^*$ ($O(p^4)$) which shows up as a
broad continuum \cite{34}.

\subsection{Decay $K^0\to$ Three Photons}
Given that $B(K_L\to 2\gamma )=5.7\times 10^{-4}$,
$B(K_S\to 2\gamma )=2.4\times 10^{-6}$ \cite{36} it is interesting
to ask what one expects for the decays $K_{L,S}\to 3\gamma $ \cite{37}.\\
\indent First of all, it should be noted that both $K_L\to 3\gamma $ and
$K_S\to 3\gamma $ are possible without violating $CP$ or any other general
symmetry principle. Gauge invariance dictates that no pair of photons in
these channels can have
$J=0$, while Bose statistics forbids any pair from having $J=1$
(Yang's theorem). It follows that every pair of photons in these
decays must have at least two units of angular momentum. A simple
model that relates the decays $K_{L,S}\to 3\gamma $ to the other quadrupole
transition $K_{L,S}\to\pi^0\pi^0\gamma $ discussed above yields \cite{37}
\begin{eqnarray}
B(K_L\to 3\gamma )&=&3\times 10^{-19},\nonumber\\
B(K_S\to 3\gamma )&=&5\times 10^{-21}.
\end{eqnarray}
Thus the $3\gamma $ decay mode is suppressed relative to the $2\gamma $
mode by {\it 15 orders of magnitude} -- a remarkable reminder of the
power of Bose statistics in this year of the Bose centenary!\\
\\
\\
\\
Acknowledgement:\\
\indent I wish to thank the organizers of the Workshop for their
affectionate hospitality, and for the opportunity to discuss physics in
a wonderfully congenial and stimulating environment.

\vspace*{2cm}
\Large
\noindent{\bf Figure Captions}
\normalsize
\begin{itemize}
\item[Fig. 1.] Diagrams relevant for the decays (a) $K^0\to
K^+e^-\bar{\nu_e}$,
(b) $K^+\to\gamma\gamma e^+\nu_e $, (c) $\Delta S=-\Delta Q$ decays $K^0\to
\pi^+l^-\bar{\nu_l}$.
\item[Fig. 2.] (a) Two--photon contribution to $K_L\to l^+l^-$ and
$K_L\to \pi^0 l^+l^-$. (b)
Diagrams describing the short--distance interaction $s\bar d\to l^+l^-$.
\item[Fig. 3.] (a) Short--distance diagrams relevant to the interaction
$s\bar d\to\nu\bar{\nu}$. (b) Long--range contributions to the decay
$K^+\to\pi^+\nu\bar{\nu}$.
\end{itemize}

\end{document}